\documentclass[12pt]{article}
\usepackage{unites2e}
\usepackage{helvet}
\usepackage{graphicx}
\usepackage{color}
\usepackage{bm}
\usepackage{amssymb}
\usepackage{amsmath}
\usepackage{epsfig}
\usepackage{wrapfig}
\usepackage{subfigure}
\usepackage{multicol}
\usepackage{url}
\begin{document}
\title{\bf \large VERITAS contributions to CF6-A: Cosmic Rays, Gamma Rays and Neutrinos}
\author{The VERITAS Collaboration}
\date{\today}             
\maketitle         
\subsubsection*{Introduction}

VERITAS, located at the Fred Lawrence Whipple Observatory in southern
Arizona, is one of three major imaging atmospheric Cherenkov telescope
facilities in operation worldwide. It consists of an array of four,
12-meter diameter telescopes, providing a $<1\%$ Crab-flux sensitivity in
the energy range between $100\U{GeV}$ and $50\U{TeV}$. The array
has been operating since 2007 and has detected 42 objects from
$\sim10$ different source classes, including many new discoveries
\cite{veritas_status}.

In 2012, the VERITAS collaboration established a long-term plan
describing a scientific strategy for operations. Many of the goals
outlined address problems relevant to the charge of the CF6
subgroup. An additional white paper describing the VERITAS indirect
dark matter detection program has been submitted to subgroup CF2
\cite{andyWP}. A prerequisite for achieving these goals was the
success of a major upgrade to the array which was completed, with no
disruption to the array operations, in the summer of 2012. The upgrade
involved the installation of a new trigger system, and the replacement
of all of the photodetectors with super-bialkali photomultiplier
tubes. This has resulted in an increase of at least 35\% in photon
collection efficiency. VERITAS will remain the premier VHE facility in
the Northern Hemisphere for some time, while the next-generation
Cherenkov Telescope Array (CTA) project is under development. The
contemporaneous overlap of VERITAS operations with Fermi-LAT, HAWC,
IceCube and Auger will be of critical importance to many of the
science goals described here.

\subsubsection*{Cosmic Particle Acceleration as Signal and Background}

VERITAS has made significant contributions to the study of Galactic
particle accelerators, including pulsars and their nebulae, gamma-ray
binary systems and supernova remnants. Highlight results include the
detection of $>100\U{GeV}$ emission from the Crab pulsar
\cite{crabpulsar}, and the first TeV detection of Tycho's SNR
\cite{tycho}. The Crab pulsar result can only be easily explained by a
new emission mechanism, or an additional component at high energies,
The Tycho detection, combined with results from the Fermi-LAT,
provides compelling evidence for hadronic particle acceleration in
SNR. Complementary evidence for a link between cosmic ray production
and star formation activity was provided by the discovery of
gamma-ray emission from the starburst galaxy M82 \cite{m82}. 

HAWC \cite{HAWC} will soon begin observations, and will provide a complete
TeV map of the northern sky. Follow-up observations with high
sensitivity and better angular and energy resolution will be performed
by VERITAS. In particular, this is key to determining the nature of
unidentified 'dark accelerators', as already demonstrated by VERITAS
observations of Milagro sources. The resolution of gamma-ray emission
in the region of MGRO~J2019+37 and the Cygnus OB1 association into at
least two distinct sources, one clearly associated with the pulsar
wind nebula CTB~87, demonstrates the importance of the excellent
angular resolution provided by the imaging technique. Contemporaneous
operation of VERITAS and HAWC will also allow a rapid response to
transient events, such as blazar flares and gamma-ray bursts.

The study of VHE gamma-ray emission from particle accelerators is, of
course, interesting from a purely astrophysical perspective. It is
also critical to understand the nature and properties of astrophysical
backgrounds in searches for new physical effects (this issue is
addressed in detail in a separate CF6 white paper
\cite{amandaWP}). The interpretation of indirect dark matter searches,
Lorentz Invariance Violation (LIV) tests, studies of gamma-ray and
antimatter backgrounds and searches for axion-like particles all rely
on an accurate knowledge of the potential astrophysical backgrounds
and their spectral, morphological and temporal properties. A classic
example of this is the case of the Galactic Center, which has the
highest local concentration of dark matter, but also hosts multiple
known and potential astrophysical TeV sources, both point-like and
extended \cite{andyWP}. VERITAS observations of the Galactic Center
are ongoing as part of our long-term observing plan. For this southern
source, the observations take place at low elevation angles, resulting
in a high energy threshold but providing an increase in the effective
collection area at high energies. This allows us to probe the end
point of the spectrum of the Galactic Center gamma-ray emission, which
may hold the key to resolving the nature of the source.


\subsubsection*{Probing Fundamental Physics}

VERITAS is a mature experiment, and has moved beyond the initial
source discovery phase. Fundamental physics topics now play an
increasingly important role in the observing plan. The success of
these studies, which often require long and technically challenging
exposures, relies on stable operation and a thorough knowledge of
the detector performance, calibration and associated Monte Carlo
simulations. After five years of operations, all of these aspects of
VERITAS are very well understood. Indirect dark matter searches are
described elsewhere \cite{andyWP}. Other topics which we plan to
investigate with VERITAS in the coming years include:

\begin{itemize}
\item{\bf Antimatter studies:} The rising positron fraction identified
  by PAMELA \cite{PAMELA} and confirmed by Fermi \cite{fermipositron,
    fermipositron2} up to a few hundred GeV is an intriguing result. It may be explained by a contribution from local astrophysical
  sources, or possibly by annihilating dark matter. First results from
  AMS-02 confirm that the positron fraction continues to rise up to at
  least $250\U{GeV}$, at which point it appears to flatten \cite{AMS}.  A
  measurement of the positron fraction at higher energies would
  provide a key discriminant between the competing
  explanations. VERITAS is attempting to make such a measurement by
  observing the shadow of the Moon in both electrons and positrons, as
  proposed by Colin \cite{colin}. This is technically challenging, due
  to the optical sky brightness close to the Moon, and the limited
  amount of observing time available at high elevations. We have
  developed short-wavelength optical filter plates for the telescope cameras
  to enable us to observe close to the Moon, and the results of
  preliminary test observations are encouraging. Observations over the
  next few years should allow us to build up the necessary exposure
  required for this unique measurement.

\item{\bf Primordial Black Holes:} In addition, or as an alternative
  to, particle dark matter, primordial black holes (PBHs) formed
  during the early universe can serve as a viable candidate for
  cosmological dark matter (see \cite{PBH}). PBHs can evaporate
  through Hawking radiation, where the evaporation rate is directly
  coupled to their mass. Consequently, during the final seconds of
  their lifetime, PBHs can release a large flux of gamma rays within
  the sensitivity range of VERITAS. Dedicated searches for these PBH
  signals have already commenced with VERITAS \cite{gordana}, and an
  evaporation rate limit of
  $\rho_{PBH}<1.29\times10^5\UU{pc}{-3}\UU{yr}{-1}$ has been placed
  using only 700 hours of VERITAS observations. This limit is already
  an order of magnitude below previous limits. VERITAS accrues
  approximately 800 hours of Moonless observations each year, so a
  significant refinement of the result can be expected.

\item{\bf Cosmological measurements using the EBL and IGMF:} The
  gamma-ray spectra of blazars are modified by interactions with
  intergalactic radiation fields through pair-production and
  subsequent cascade processes. As a result, these spectra contain an
  imprint of the extragalactic background light (EBL) and the
  intergalactic magnetic field (IGMF). The EBL comprises the combined
  flux of all extragalactic sources integrated over the history of the
  Universe, and carries unique information regarding the epoch of
  galaxy formation and the history of galaxy evolution. This topic is
  discussed in detail in a related white paper \cite{frankWP}. The
  IGMF strength is only weakly constrained, and impossible to measure
  directly. VERITAS observations of the spectra, angular distribution
  and arrival times of gamma-rays from distant blazars will provide
  constraints to, or a measurement of, the IGMF strength which is not
  accesible to other techniques. A positive measurement would be
  important, possibly implying the existence of a primordial
  field produced in the early Universe. Both EBL and IGMF measurements
  require deep, multi-year exposures of numerous blazars over a range
  of redshifts out to $z\sim0.5$, as envisaged in our long-term
  observing plan.

\item{\bf Tests of Lorentz Invariance Violation (LIV):} Blazar observations
  provide the most stringent tests of LIV for VERITAS, thanks to their
  large distance and rapid timescale of variability. Four bright,
  high-energy peaked BL Lac objects have been identified for deep
  monitoring exposures of $\geq100\U{hours}$ in our long-term plan. An
  additional target-of-opportunity program allows us to respond
  rapidly to alerts of enhanced emission from instruments at other
  wavelengths. The detection of VHE emission from the Crab pulsar also
  raises the possibility of using pulsar time profiles to constrain LIV
  \cite{crabLIV}, and we plan to substantially augment our already
  extensive Crab pulsar dataset over the coming years, as well as to
  search for pulsed emission from other candidate sources. The energy
  threshold reduction provided by the 2012 upgrade will be
  particularly important in this regard.

\end{itemize}

\subsubsection*{UHECRs and Neutrino Astrophysics}

VERITAS observations impact the related fields of ultra-high energy
cosmic rays (UHECRs) and neutrino astrophysics. The UHECRs are most
likely extragalactic in origin, with active galactic nuclei (AGN)
among the best candidates for the accelerators. Gamma-ray observations
in the GeV-TeV range are essential to constrain models of particle
acceleration and gamma-ray/ neutrino emission in these sources (see
\cite{dummWP} for more details). Our long-term plan calls for regular
monitoring of most of the northern hemisphere VHE blazar population
over the next five years, allowing us to accumulate deep exposures of
the sources in various emission states, and maximizing our chances of
detecting bright VHE flares. Observations of the nearby radio galaxy
M87 will also continue, and will be complemented by high resolution
X-ray and radio observations in the event of a flare. A clear
correlation between morphological changes in the jet structure and the
VHE emission state could help to finally pin down the particle
acceleration and photon emission region in AGN jets.

VERITAS can also act as a flare alert system for the UHECR and
neutrino observatories, and provide rapid, high sensitivity follow-up
observations. In response to the early Auger reports of a correlation
between ultra-high energy cosmic rays and AGN, VERITAS was the first
instrument to provide follow-up TeV gamma-ray observations
\cite{UHECRs}. No gamma-ray emission was seen, and the evidence for a
correlation has diminished over time, but the observations demonstrate
the substantial overlap between the two instruments, despite their
locations in different hemispheres. IceCube, conversely, can easily
view the northern sky, and VERITAS and IceCube are very well-matched
in energy range (IceCube has a minimum neutrino energy threshold of
50-100GeV, and an optimal response above $1\U{TeV}$
\cite{IceCube}). Numerous predictions of measurable neutrino fluxes
associated with astrophysical particle accelerators exist in the
literature, including both Galactic (SNRs, binary systems,
unidentified TeV sources and pulsar wind nebulae \cite{neutrinos_gal1,
  neutrinos_gal2}) and extragalactic (GRBs, active and starburst
galaxies \cite{neutrinos_xgal}) objects. VERITAS is the best instrument
to search for and characterize the electromagnetic signatures of
particle interactions in these objects, which will be necessary to
assess the relative contributions of leptonic and hadronic particle
populations. We will perform follow-up gamma-ray observations of any
reported neutrino sources, and have established a memorandum of
understanding between VERITAS and IceCube which allows us to rapidly
trigger observations of any transient neutrino excess. IceCube and
Auger will be at their most productive over the coming five years, and
VERITAS observations will both complement and augment their results.


\end{document}